\documentclass[16pt]{article}
\begin{document}

\title{\textbf{\textsf{Interacting dark energy with inhomogeneous equation of state}}}
\author{ Mubasher Jamil\footnote{Corresponding author: mjamil@camp.edu.pk}\ \ and Muneer Ahmad Rashid\footnote{muneerrshd@yahoo.com}
\\ \\
\textit{Center for Advanced Mathematics and Physics}\\
\textit{National University of Sciences and Technology}\\
\textit{Peshawar Road, Rawalpindi, 46000, Pakistan} \\
} \maketitle
\begin{abstract}
We have investigated the model of dark energy interacting with dark
matter by choosing inhomogeneous equations of state for the dark
energy and a non-linear interaction term for the underlying
interaction. The equations of state have dependencies either on the
energy densities, the redshift, the Hubble parameter or the bulk
viscosity. We have considered these possibilities and have derived
the effective equations of state for the dark energy in each case.

\end{abstract}

\section{Introduction}

One of the outstanding developments in astrophysics in the past
decade is the discovery that the expansion of the universe is
accelerated, supposedly driven by some exotic vacuum energy
\cite{perl,ries,sper,sper1,cope}. Surprisingly, the energy density
of the vacuum energy is two-third of the critical density (
$\Omega_\Lambda\simeq0.7$) apart from dark matter
($\Omega_m\simeq0.3$). The astrophysical data suggest that this
change in the expansion history of the universe is marginally recent
($z\simeq0.7$) compared with the age of the universe. The nature and
composition of dark energy is still unresolved, but by using
thermodynamical considerations, it is conjectured that the
constituents of dark energy may be massless particles (bosons or
fermions) whose collective behavior resembles a kind of radiation
fluid with negative pressure. Moreover, the temperature of the
universe filled with dark energy will increase as the universe
expands \cite{lima}. The earliest proposal to explain the recent
accelerated expansion was the cosmological constant $\Lambda$
represented by the equation of state (EoS) $p=-\rho$ (or
$\omega=-1$) having a negative pressure. In order to comply with the
data, the cosmological constant has to be fine tuned up to 56 to 120
orders of magnitude \cite{doglov}, which requires extreme fine
tuning of several cosmological parameters. It also posed the
coincidence problem in cosmology (the question of explaining why the
vacuum energy came to dominate the universe very recently)
\cite{bento1}. This latter problem is addressed through the notion
of a tracker field $Q$, in which the tracker field rolls down a
potential $V(Q)$ according to an attractor-like solution to the
equations of motion \cite{zlat}. But here the field has difficulties
reaching $\omega<-0.7$, while current observations favor
$\omega<-0.78$ with 95\% confidence level \cite{linder}. It is shown
that a quintessence scalar field coupled with either a dissipative
matter field, a Chaplygin gas (CG) or a tachyonic fluid solves the
coincidence problem \cite{chim}. These problems are alternatively
discussed using anthropic principles as well \cite{wein}. Several
other models have been proposed to explain the cosmic accelerated
expansion by introducing decaying vacuum energy \cite{free,frie}, a
cardassian term in the Friedmann-Robertson-Walker (FRW) equations
\cite{free1}, a generalized Chaplygin gas (GCG) \cite{bento} and a
phantom energy ($\omega<-1$) arising from the violation of energy
conditions \cite{cald,babi,ness}. Another possibility is the
`geometric dark energy' based on the Ricci scalar $R$ represented by
$\Re=R/12H^2$, where $H$ is the Hubble parameter \cite{linder}.
Notice that $\Re>1/2$ represents accelerated expansion, and $\Re>1$
gives a super-accelerated expansion of the universe, whereas
presently $\Re=1/2$.

Models based on dark energy interacting with dark matter have been
widely investigated
\cite{vage,sami,lin,wu,wang1,jamil,jamil1,zim,seta,set1,rub,mota}.
These models yield stable scaling solution of the FRW equations at
late times of the evolving universe. Moreover, the interacting CG
allows the universe to cross the phantom divide (the transition from
$\omega>-1$ to $\omega<-1$), which is not permissible in pure CG
models. In fact it is pointed out that a phantom divide (or
crossing) is possible only if the cosmic fluids have some
interaction \cite{vik}. It is possible that this interaction can
arise from the time variation of the mass of dark matter particles
\cite{zhang1}. It is shown that the cosmic coincidence problem is
fairly alleviated in the interacting CG models \cite{campo}. This
result has been endorsed with interacting dark energy in \cite{sad}.
There is a report that this interaction is physically observed in
the Abell cluster A586, which in fact supports the GCG cosmological
model and apparently rules out the $\Lambda$CDM model
\cite{bertolami}. However, a different investigation of the
observational $H(z)$ data rules out the occurrence of any such
interaction and favors the possibility of either more exotic
couplings or no interaction at all \cite{wei}. In this context, we
have investigated the interaction of the dark energy with dark
matter by using a more general interaction term. We have focused on
the inhomogeneous EoS for dark energy as these are
phenomenologically relevant.

The outline of the paper is as follows. In the next section, we
present a general interacting model for our dynamical system. In the
third section, we derive the effective EoS for the interacting dark
energy by employing different inhomogeneous EoS, having dependencies
on various cosmological parameters. Finally, we present our
conclusion.

\section{The interacting model}

We assume the background to be a spatially homogeneous and isotropic
FRW spacetime, given by
\begin{equation}
ds^2=-dt^2+a^2(t)\left[\frac{dr^2}{1-kr^2}+r^2(d\theta^2+\sin^2\theta
d\phi^2)\right],
\end{equation}
filled with the two component fluid namely dark energy and dark
matter. Here $a(t)$ is the scale factor and $k=-1,0,1$ represents
the spatially hyperbolic, flat or closed universe, respectively. The
corresponding Einstein field equation is
\begin{equation}
H^2\equiv\left(\frac{\dot{a}}{a}\right)^2=\frac{\kappa}{3}\rho-\frac{k}{a^2},
\end{equation}
where $\kappa=8\pi G$ and $\rho=\rho_\Lambda+\rho_m$. Moreover, the
energy conservation for our gravitational system is given by
\begin{equation}
\dot{\rho}+3H(\rho+p)=0,
\end{equation}
where $p=p_\Lambda=\omega_\Lambda \rho_\Lambda$ and $p_m=0$ or
$\omega_m=0$. We assume a special form of the interaction
$Q=\Gamma\rho_\Lambda$ between dark energy and dark matter, where
$\Gamma$ is the decay rate. Then Eq. (3) can be subdivided into two
parts, corresponding to $\rho_\Lambda$ and $\rho_m$ as follows:
\begin{eqnarray}
\dot{\rho}_{\Lambda}+3H(1+\omega_\Lambda)\rho_\Lambda &=&-Q,\\
\dot{\rho}_m+3H\rho_m &=&Q,
\end{eqnarray}
respectively. Eqs. (4) and (5) show that the energy conservation for
dark energy and matter would not hold independently if there is
interaction between them but would hold globally for the whole
interacting system as is manifest in Eq. (3). We further define the
density ratio $r_m$, by a form of scaling relation, by
$r_m\equiv\rho_m/\rho_\Lambda$. To study how this density ratio
evolves with time, we differentiate $r_m$ with respect to $t$:
\begin{equation}
\dot{r}_m=\frac{dr_m}{dt}=\frac{\rho_{m}}{\rho_{\Lambda}}\left[\frac{\dot{\rho}_{m}}{\rho_{m}}-\frac{\dot{\rho}_{\Lambda}}{\rho_{\Lambda}}\right].
\end{equation}
Using Eqs. (4) and (5) in (6), we get
\begin{equation}
\dot{r}_m=3Hr_m\left[\omega_\Lambda+\frac{1+r_m}{r_m}\frac{\Gamma}{3H}\right].
\end{equation}
Furthermore, we define an effective EoS for dark energy and matter
by \cite{kim}
\begin{equation}
\omega_\Lambda^{eff}=\omega_\Lambda+\frac{\Gamma}{3H}, \ \
\omega_m^{eff}=\frac{-1}{r_m}\frac{\Gamma}{3H},
\end{equation}
which also involve the contribution from the interaction between
matter and dark energy. Using Eq. (8) in Eqs. (4) and (5), we get
\begin{equation}
\dot{\rho}_{\Lambda}+3H(1+\omega_\Lambda^{eff})\rho_\Lambda=0,
\end{equation}
\begin{equation}
\dot{\rho}_m+3H(1+\omega_m^{eff})\rho_m=0.
\end{equation}
From the standard FRW model, the density parameters corresponding to
matter and dark energy are defined by
\begin{equation}
\Omega_m=\frac{\rho_m}{\rho_{cr}},\ \
\Omega_\Lambda=\frac{\rho_\Lambda}{\rho_{cr}}.
\end{equation}
The above parameters are related by $\Omega_m+\Omega_\Lambda=1.$
Using the definition of $r_m$, we can write
\begin{equation}
r_m\equiv\frac{\Omega_m}{\Omega_\Lambda}=\frac{1-\Omega_\Lambda}{\Omega_\Lambda}.
\end{equation}
The value of $r_m$ decreases monotonically with expansion and varies
very slowly at the present era. Contrary to the noninteracting case,
$r_m$ decreases slower when there is an interaction \cite{wang3}.
Using the definition of $r_m$ in Eq. (2), we get
\begin{equation}
H^2=\frac{\kappa}{3}(1+r_m)\rho_\Lambda.
\end{equation}
Next, we choose the following generalized interaction term
\begin{equation}
Q=3Hc[\gamma\rho_m+\beta\rho_\Lambda+\delta(\rho_m\rho_\Lambda)^{1/2}]^n,
\end{equation}
with the corresponding decay rate
\begin{equation}
\Gamma=3Hc(\beta+\gamma r_m+\delta\sqrt{r_m})^n,
\end{equation}
which follows from $Q=\Gamma\rho_\Lambda^n$ where $n$, $\beta$,
$\gamma$ and $\delta$ are constant parameters. The above-mentioned
$c$ is the coupling constant. Notice that $c>0$ yields conversion of
dark energy into dark matter, and vice versa if $c<0$. Note that for
$\beta=\gamma=n=1$ and $\delta=0$, Eq. (14) reduces to the usual
linear interaction term \cite{wang2}. Making use of Eq. (14) in (4),
the EoS parameter becomes
\begin{equation}
\omega_\Lambda=-1-\frac{\dot{\rho}_\Lambda}{3H\rho_\Lambda}-\frac{\Gamma
\rho_\Lambda^{n-1}}{3H}.
\end{equation}
Using Eq. (16) in (8), the effective EoS of the dark energy is given
by
\begin{equation}
\omega_\Lambda^{eff}=-1-\frac{\dot{\rho}_\Lambda}{3H\rho_\Lambda}+\frac{\Gamma}{3H}(1-\rho_\Lambda^{n-1}).
\end{equation}
In the forthcoming discussion, we shall determine the effective EoS
for dark energy corresponding to various equations of state.

\section{Inhomogeneous equations of state for dark energy}

The general EoS relating the pressure density $p$ and energy density
$\rho$ is given by
\begin{equation}
F(p_\Lambda,\rho_\Lambda)=0
\end{equation}
We shall also consider equations of state depending on either the
redshift $z$, the scale factor $a(t)$ or the bulk viscosity $\xi$.
Note that we are not considering an EoS explicitly containing the
time $t$, as we always have the opportunity to use $a(t)=\varepsilon
t^{-\lambda}$, with $\varepsilon$ and $\lambda$ constant parameters.

\subsection{Generalized cosmic Chaplygin gas}

We now take the EoS of the generalized cosmic Chaplygin gas given by
\cite{diaz}
\begin{equation}
p_\Lambda=-\rho_\Lambda^{-\alpha}[C+(\rho_\Lambda^{1+\alpha}-C)^{-\sigma}],
\end{equation}
where $C=\frac{A}{1+\sigma}-1$ with $\alpha>1$ and $A$ constant
parameters and $-l<\sigma<0$, where $l>1$. This EoS reduces to that
of the generalized Chaplygin gas if furthermore $\sigma=0$ and to
the Chaplygin gas if further $\alpha=1$. The motivation to use this
EoS is to construct the cosmological models that are stable and free
from nonphysical behaviors even when the vacuum fluid behaves like a
phantom energy \cite{writ}.

Using the energy conservation principle, the density evolution is
\begin{equation}
\rho_\Lambda=[C+(1+C_1a^{-3(1+\alpha)(1+\sigma)})^{\frac{1}{1+\sigma}}]^{\frac{1}{1+\alpha}},
\end{equation}
where $C_1$ is the constant of integration. We define
\begin{equation}
\Delta_1\equiv(\rho_\Lambda^{1+\alpha}-C)^{1+\sigma}-1=C_1a^{-3(1+\alpha)(1+\sigma)}.
\end{equation}
Making use of Eqs. (20) and (21) in (17), the effective EoS for dark
energy becomes
\begin{equation}
\omega_\Lambda^{eff}=-1+\frac{\Delta_1(1+\Delta_1)^{\frac{-\sigma}{1+\sigma}}}{C+(1+\Delta_1)^\frac{1}{1+\sigma}}+
\frac{\Gamma}{3H}(1-\rho_\Lambda^{n-1}),
\end{equation}
where $\rho_\Lambda$ is determined by Eq. (20).

\subsection{New generalized Chaplygin gas}

Zhang et al. \cite{zhang1} suggested another general form of
Chaplygin gas, called the new generalized Chaplygin gas given by
\begin{equation}
p_\Lambda=\frac{-\tilde{A}(a)}{\rho_\Lambda^\alpha},\ \
\tilde{A}(a)=-\omega_\Lambda Aa^{-3(1+\omega_\Lambda)(1+\alpha)}.
\end{equation}
Here $\alpha$ is a constant parameter. This model is dual to the
interacting XCDM model, where the X part corresponds to quintessence
or X-matter ($\omega_\Lambda<-1/3$).

In this model, the energy density  evolves as
\begin{equation}
\rho_\Lambda=[Aa^{-3(1+\omega_\Lambda)(1+\alpha)}+C_2a^{-3(1+\alpha)}]^{1/(1+\alpha)},
\end{equation}
where $C_2$ is a constant of integration. Thus using Eq. (24) in
(17) the effective EoS is given by
\begin{equation}
\omega_\Lambda^{eff}=-1+\frac{\omega_\Lambda+\Delta_2}{\Delta_2}+\frac{\Gamma}{3H}(1-\rho_\Lambda^{n-1}),
\end{equation}
where
\begin{equation}
\Delta_2\equiv A+C_2a^{3\omega_\Lambda(1+\alpha)},
\end{equation}
and $\rho_\Lambda$ is determined from Eq. (24).

\subsection{Generalizing the generalized Chaplygin gas}

Sen and Scherrer \cite{sen} suggested an EoS for the generalized
Chaplygin gas by assuming the constant parameter $\alpha$ to be
free, where we have
\begin{equation}
\omega_\Lambda=-\frac{A_s}{A_s+(1-A_s)(\frac{a}{a_o})^{-3(1+\alpha)}},
\end{equation}
where
\begin{equation}
A_s=\frac{A}{\rho_{\Lambda_o}^{1+\alpha}}.
\end{equation}
Using Eq. (27) one can have various cosmological scenarios: for
$0<A_s<1$ and $\alpha>-1$ we have the standard generalized Chaplygin
gas model giving dark matter-dark energy unification. For $A_s>1$
and $\alpha>-1$, it gives the early phantom generalized Chaplygin
gas, i.e., it behaves as phantom energy at early times and behaves
like the cosmological constant $\omega_\Lambda=-1$ at late times.
For $0<A_s<1$ and $\alpha<-1$ it represents the transient
generalized Chaplygin gas, in which case Eq. (27) gives de Sitter
regime at early times and a matter dominated regime at later times.

The density evolution is given by
\begin{equation}
\rho_\Lambda=\rho_{\Lambda_o}\left[A_s+(1-A_s)(\frac{a}{a_o})^{-3(1+\alpha)}\right]^{1/1+\alpha}.
\end{equation}
The corresponding effective EoS is
\begin{equation}
\omega_\Lambda^{eff}=-A_s\left(\frac{\rho_\Lambda}{\rho_{\Lambda_o}}\right)^{-(1+\alpha)}+\frac{\Gamma}{3H}(1-\rho_{\Lambda}^{n-1}),
\end{equation}
with $\rho_\Lambda$ determined by Eq. (29).

\subsection{Interacting scale factor dependent dark energy}

We here take the EoS \cite{sri}
\begin{equation}
p_\Lambda=-\rho_\Lambda(1+A a^\alpha).
\end{equation}
The corresponding density evolution is
\begin{equation}
\rho_\Lambda=C_3\exp{\left(\frac{3Aa^\alpha}{\alpha}\right)},
\end{equation}
with $C_3$ is constant of integration. The effective EoS is given by
\begin{equation}
\omega_\Lambda^{eff}=-Aa^\alpha+\frac{\Gamma}{3H}(1-\rho_{\Lambda}^{n-1}).
\end{equation}

\subsection{Interacting Hubble parameter dependent dark energy}

An interesting EoS depending on the Hubble parameter $H$ is given by
\cite{noji}
\begin{equation}
p_{\Lambda}=-\rho_{\Lambda}+f(\rho_{\Lambda})+G(H).
\end{equation}
The corresponding FRW equation is
\begin{equation}
\dot{\rho}_{\Lambda}=-3H[f(\rho_{\Lambda})+G(H)].
\end{equation}
Let us choose the following EoS:
\begin{equation}
f(\rho_{\Lambda})+G(H)=-A\rho_{\Lambda}^\alpha-BH^{2\epsilon},
\end{equation}
where $ \epsilon$ is a constant. Using Eq. (13) in (36), we get
\begin{equation}
f(\rho_{\Lambda})+G(H)=-A\rho_\Lambda^\alpha-B^{\prime}\rho_\Lambda^\epsilon,
\end{equation}
where
\begin{equation}
B^{\prime}\equiv B\left[\frac{\kappa}{3}(1+r_m)\right]^\epsilon.
\end{equation}
Using Eqs. (35) and (36) in (17), we get
\begin{equation}
\omega_\Lambda^{eff}=-1-(A\rho_\Lambda^{\alpha-1}+B^{\prime}\rho_\Lambda^{\epsilon-1})+\frac{\Gamma}{3H}(1-\rho_\Lambda^{n-1}).
\end{equation}
Here $\rho_\Lambda$ is determined from Eq. (13).

\subsection{Interacting redshift dependent dark energy}

We here assume that the dark energy evolves with the redshift
parameter $z$. Hence we take the following linear EoS \cite{upa1}:
\begin{equation}
\omega(z)=\omega_o+\omega_1z,
\end{equation}
where $\omega_o$ and $\omega_1$ are constants. This EoS was used to
analyze the cosmic microwave background and the matter power
spectrum \cite{upa}. Eq. (40) effectively works for $z<1$, while
$\omega(z)=\omega_o+\omega_1$ holds up till $z\approx1$. The
thermodynamical properties of dark energy have been investigated
using Eq. (40), and it is deduced that the apparent horizon of the
universe may be the boundary of thermodynamical equilibrium for the
universe like the event horizon for a black hole \cite{zhang2}. We
are interested in the evolution of dark energy (i.e. Eq. (40)) in
our generalized interacting model. The energy conservation principle
gives
\begin{equation}
\rho_\Lambda=C_3a^{-3(1+\omega_o-\omega_1)}\exp{\left(3\omega_1\frac{a_o}{a}\right)},
\end{equation}
where we have used $z\equiv(a_o/a)-1$ and $C_3$ is a constant of
integration. Notice that for $\omega_1=0$, Eq. (41) gives the
evolution of the usual dark energy.

Using Eq. (41) in (17), we get
\begin{equation}
\omega_\Lambda^{eff}=\omega_o+\omega_1z+\frac{\Gamma}{3H}(1-\rho_{\Lambda}^{n-1}).
\end{equation}
The astrophysical data support cosmologies with
$\omega_o=-1.25\pm0.09$ and $\omega_1=1.97^{+0.08}_{-0.01}$
\cite{zhang2}. Also using $0\leq c\leq1$, we see that the right hand
side of Eq. (42) becomes negative, i.e. $\omega_\Lambda^{eff}<0$,
thereby supporting the existence of phantom energy.

We now take another EoS \cite{linder,nojiri}:
\begin{equation}
\omega(z)=\omega_o+\omega_1\left(1-\frac{a}{a_o}\right)=\omega_o+\omega_1\left(\frac{z}{1+z}\right).
\end{equation}
It avoids the divergent behavior as opposed to Eq. (40) and hence is
used to parameterize the astrophysical data to higher redshift, up
till $z\approx z_{rec}$.  The two constants appearing in Eq. (43)
are constrained: $-1\leq\omega_o\leq-0.434$ and
$-0.564\leq\omega_1\leq0.498$ \cite{huang}. Its density evolution is
given by
\begin{equation}
\rho_\Lambda=C_4a^{-3(1+\omega_o+\omega_1)}\exp{\left(3\omega_1\frac{a}{a_o}\right)},
\end{equation}
where $C_4$ is a constant of integration. Thus the effective EoS is
\begin{equation}
\omega_\Lambda^{eff}=\omega_o+\omega_1\left(1-3\frac{a}{a_o}\right)+\frac{\Gamma}{3H}(1-\rho_{\Lambda}^{n-1}),
\end{equation}
with $\rho_\Lambda$ is determined from Eq. (44).

\subsection{Interacting viscous dark energy}

The Eckart theory \cite{ecka} effectively deals with fluids having
non-zero viscosities. The term viscosity arises from fluid dynamics,
which has two major parts, namely the bulk viscosity $\xi$ and the
shear viscosity $\eta$. In viscous cosmology, shear viscosities
arise in relation to space anisotropy, while the bulk viscosity
accounts for the space isotropy \cite{brevik,hu,coles}. The bulk
viscous fluid is represented by the Eckart expression $\Pi=-\xi
(\rho_\Lambda )u_{;\mu }^{\mu }$, where $u^\mu$ is the four velocity
of the viscous fluid. The bulk viscosity is generally taken to be
positive to ensure positive entropy production in conformity with
the second law of thermodynamics \cite{Zim}. Its scaling may be
represented by $\xi \sim \rho_\Lambda ^{-\zeta }$, where $\zeta$ is
a constant parameter.

The effective pressure containing the isotropic pressure and the
viscous stress is given by
\begin{equation}
p_{eff}=p_\Lambda+\Pi,
\end{equation}
where $p_\Lambda=\chi/\rho_\Lambda^\alpha$ with $\chi>0$ and
$\Pi=-3H\xi$ in the FRW model. Thus the energy conservation
principle for the bulk viscous fluid becomes
\begin{equation}
\dot{\rho}_\Lambda+3H(\rho_\Lambda+p_\Lambda-3H\xi)=0.
\end{equation}
Using Eq. (13) in (47) we have
\begin{equation}
\dot{\rho}_\Lambda+3H(\rho_\Lambda+p_\Lambda-\xi\Upsilon\sqrt{\rho_\Lambda})=0,
\end{equation}
where $\Upsilon\equiv\sqrt{3\kappa(1+r_m)} $. Solving Eq. (48) we
get
\begin{equation}
a(t)=\left( \frac{1}{C_5}\exp{\left[\int\frac{\rho_\Lambda^\alpha
d\rho_\Lambda}{\rho_\Lambda^{1+\alpha}-\xi\Upsilon\rho_\Lambda^{\alpha+\frac{1}{2}}+\chi}\right]}
\right)^{-1/3},
\end{equation}
with $C_5$ is a constant of integration. This equation can be solved
exactly by choosing $\xi=\upsilon\rho_\Lambda^{1/2}$ with $\upsilon$
a constant; thus, we have
\begin{equation}
\rho_\Lambda=\left[\frac{(C_5a)^{-3(1-\upsilon\Upsilon)(1+\alpha)}-\chi}{1-\upsilon\Upsilon}\right]^{\frac{1}{1+\alpha}}.
\end{equation}
Defining
\begin{equation}
\Delta_3\equiv(C_5a)^{-3(1-\upsilon\Upsilon)(1+\alpha)}=\rho_\Lambda^{1+\alpha}(1-\upsilon\Upsilon)+\chi.
\end{equation}
Using Eqs. (50) and (51) in (17), we get the effective EoS of the
interacting viscous dark energy:
\begin{equation}
\omega_\Lambda^{eff}=-1+(1-\upsilon\Upsilon)\left(\frac{\Delta_3}{\Delta_3-\chi}\right)+\frac{\Gamma}{3H}(1-\rho_{\Lambda}^{n-1}).
\end{equation}

\section{Conclusion}
In this work, we have determined various effective equations of
state for the dark energy having non-zero interaction with the
matter in the universe. The dark energy can have dependencies on
various cosmological parameters like the Hubble parameter, the
redshift, the scale factor, the energy densities or the bulk
viscosity. We have considered all such possibilities in our
interacting model.

\subsection*{Acknowledgments}
We would like to thank X. Zhang and S. Odintsov for useful
discussions during this work.

\end{document}